\documentclass[10pt,letterpaper,twocolumn]{article}
\usepackage{ol2}
\usepackage[draft]{hyperref}
\usepackage{amsmath}

\begin{document}

\twocolumn[
\title{Stability of solitons in $\mathcal{PT}$-symmetric couplers}
\author{Rodislav Driben,$^{1,2}$ and Boris A. Malomed$^{2}$}
\address{
$^1$Jerusalem College of Engineering - Ramat Beit HaKerem,
POB 3566, Jerusalem, 91035, Israel \\
$^2$Department of Physical Electronics, School of Electrical Engineering,
Faculty of Engineering, Tel Aviv University, Tel Aviv 69978, Israel}

\begin{abstract}Families of analytical solutions are found for symmetric and
antisymmetric solitons in the dual-core system with the
Kerr nonlinearity and $\mathcal{PT}$-balanced gain and loss. The
crucial issue is stability of the solitons. A stability
region is obtained in an analytical form, and verified by
simulations, for the $\mathcal{PT}$-symmetric solitons. For the
antisymmetric ones, the stability border is found in a numerical
form. Moving solitons of both types collide elastically. The
two soliton species merge into one in the ``supersymmetric"case,
with equal coefficients of the gain, loss and inter-core coupling.
These solitons feature a subexponential instability, which may be
suppressed by periodic switching (``management").\end{abstract}

\ocis{060.5530, 190.6135, 190.5940, 230.4320}

 ] 

Dissipative media featuring the parity-time ($\mathcal{PT}$) symmetry have
recently drawn a great deal of attention. The introduction of this symmetry
in optics followed works extending the canonical quantum theory~to
non-Hermitian Hamiltonians that may exhibit a real spectrum~\cite{Bender}.
The Hamiltonian is $\mathcal{PT}$-symmetric if it includes a complex
potential $V(x)$ which satisfies constraint $V(x)=V^{\ast }(-x)$. Such
potentials were proposed \cite{Muga}-\cite{DC-review} and realized \cite%
{experiment1,Moti} in optics, by juxtaposing spatially symmetric patterns of
the refractive index and appropriately placed gain and loss~elements, see
Ref. \cite{DC-review} for a review. Nonlinear $\mathcal{PT}$ systems \cite%
{nonlin2}-\cite{Ufa} and respective solitons \cite{soliton} were introduced
too.

A medium which is akin to $\mathcal{PT}$ systems is a dual-core
waveguide with gain and loss acting separately in two cores, which
are linearly coupled by the tunneling of light \cite{Herb}. This
system predicts stable 1D solitons in optical
\cite{Herb}-\cite{Pavel} and plasmonic \cite{Dima} waveguides with
the Kerr nonlinearity, as well as 2D dissipative solitons and
vortices \cite{vort,vort2}. The system is made $\mathcal{PT}$
symmetric by adopting equal strengths of the gain and loss in the
cores. A challenging problem is the stability of solitons, as stable
pulses in the dual-core system were previously found far from the
point of the $\mathcal{PT}$ symmetry \cite{Javid,Pavel,Dima}. We
produce two families of exact soliton solutions for the
$\mathcal{PT}$-symmetric system, which correspond to symmetric and
antisymmetric solitons in the ordinary dual-core coupler
\cite{Wabnitz}-\cite{Pak}. For the former family, an exact stability
border is found analytically, and verified by simulations. For the $\mathcal{%
PT}$-antisymmetric solitons, the stability region is identified in a
numerical form. In the ``supersymmetric" limit, when the gain and
loss coefficients coincide with the constant of the inter-core
coupling, both families merge into solitons which are subject to a
subexponential instability. We demonstrate that this instability can
be suppressed by means of a ``management" technique, i.e., periodic
switching of the gain, loss, and coupling.



The transmission of light or plasmons in the dual-core waveguide is
described by the linearly coupled equations for amplitudes $u(z,t)$ and $%
v(z,t)$ in the active and passive cores \cite{Herb}-\cite{Dima}:%
\begin{eqnarray}
iu_{z}+(1/2)u_{tt}+|u|^{2}u-i\gamma u+\kappa v &=&0,  \notag \\
iv_{z}+(1/2)v_{tt}+|v|^{2}v+i\Gamma v+\kappa u &=&0,  \label{uv}
\end{eqnarray}%
where $z$ is the propagation distance and $t$ the reduced time or transverse
coordinate in the temporal- or spatial-domain system. Coefficients
accounting for the dispersion or diffraction, Kerr nonlinearity, and
inter-core coupling, $\kappa $, are normalized to be $1$, while $\gamma $
and $\Gamma $ are coefficients of the linear gain and loss in the two cores.
The $\mathcal{PT}$ symmetry in Eqs. (\ref{uv}) corresponds to $\Gamma
=\gamma $. Strictly speaking, the $\mathcal{PT}$-balanced gain and loss
correspond to the border between stable and unstable systems: The zero
solution, $u=v=0$, is unstable at $\gamma >\Gamma $, while the stability
region was found at $\gamma <\Gamma $ \cite{Herb,Javid}.

Obviously, \emph{any} solution to the nonlinear Schr\"{o}dinger (NLS)
equation (with a frequency shift), $iU_{z}+(1/2)U_{tt}+|U|^{2}U\pm \sqrt{%
1-\gamma ^{2}}U=0,$ gives rise to two \emph{exact} solutions of the $%
\mathcal{PT}$-symmetric system, provided that $\gamma \leq 1$:
\begin{equation}
v=\left( i\gamma \pm \sqrt{1-\gamma ^{2}}\right) u=U\left( z,t\right)
\label{U}
\end{equation}%
(recall we fix $\kappa \equiv 1$). For $\gamma =0$, solutions (\ref{U}) with
$+$ and $-$ amount, respectively, to the symmetric and antisymmetric modes
in the dual-core coupler \cite{Wabnitz}-\cite{Pak}, therefore we call the
respective solutions (\ref{U}) $\mathcal{PT}$-symmetric and $\mathcal{PT}$%
-antisymmetric ones. In the limit of $\gamma =1$, exact solutions (\ref{U})
reduce to a single one, $v=iu=U(z,t)$. In particular, $\mathcal{PT}$%
-symmetric and antisymmetric solitons, with arbitrary amplitude $\eta $, are
generated by the NLS solitons,%
\begin{equation}
U\left( z,t\right) =\eta \exp \left( i\left( \eta ^{2}/2\pm \sqrt{1-\gamma
^{2}}\right) z\right) \mathrm{sech}\left( \eta t\right) .  \label{sol}
\end{equation}

As concerns stability of the solitons, it is again relevant to compare the $%
\mathcal{PT}$-symmetric system to its counterpart with $\gamma =0$. The
exact result for the latter system is that the symmetric solitons are
unstable against symmetry-breaking perturbations at $\eta ^{2}>\eta _{\max
}^{2}\left( \gamma =0\right) \equiv 4/3$ \cite{Wabnitz}. The antisymmetric
solitons are unstable against oscillatory perturbations at $\eta ^{2}>\tilde{%
\eta}_{\max }^{2}\approx 0.8$ \cite{Akhmed}. The latter instability
threshold was found in a numerical form.

The analysis of the instability against antisymmetric perturbations, $\delta
u=-\delta v$, which leads to the exact result $\eta _{\max }^{2}\left(
\gamma =0\right) \equiv 4/3$, can be extended for the $\mathcal{PT}$%
-symmetric solitons. The respective perturbation $\delta u$ at the critical
point, $\eta ^{2}=\eta _{\max }^{2}$, obeys the linearized equation, $%
\left\{ 4\sqrt{1-\gamma ^{2}}-d^{2}/dt^{2}+\eta _{\max }^{2}\left[ 1-6%
\mathrm{sech}^{2}\left( \eta _{\max }t\right) \right] \right\} \delta u=0.$
This solvable equation yields
\begin{equation}
\eta _{\max }^{2}\left( \gamma \right) =(4/3)\sqrt{1-\gamma ^{2}}.
\label{crit}
\end{equation}

\begin{figure}[tbh]
\centerline{\includegraphics[width=8.0cm]{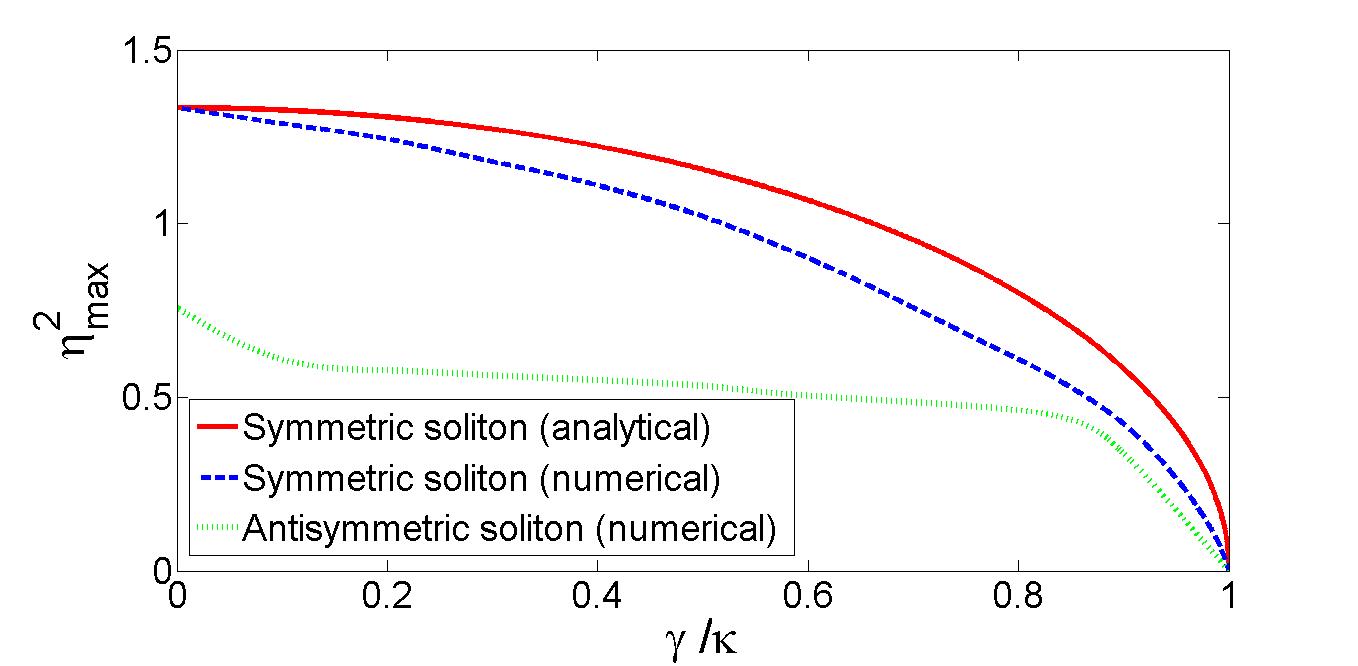}}
\caption{(Color online) The analytical stability border (\protect\ref{crit})
and its numerically found counterpart for the $\mathcal{PT}$-symmetric
solitons, and the numerical stability border for the $\mathcal{PT}$%
-antisymmetric solitons. The solitons with amplitude $\protect\eta $ are
stable at $\protect\eta <\protect\eta _{\max }$.}
\label{fig1}
\end{figure}

We have performed simulations of the evolution of perturbed solitons within
the framework of Eqs. (\ref{uv}), aiming to verify the stability border (\ref%
{crit}) of the $\mathcal{PT}$-symmetric solitons, and identify a border for
their antisymmetric counterparts. Perturbations were introduced by
independently changing the initial amplitudes of both components by $\pm 3\%$%
. The results are summarized in Fig. \ref{fig1} for both soliton families.
The numerical stability border for the symmetric solitons goes somewhat
below the analytical one (\ref{crit}) because the finite perturbations used
in the simulations are actually rather strong. Accordingly, taking smaller
perturbations, we obtain the numerical stability border approaching the
analytical limit. For instance, at $\gamma =0.5$, amplitude perturbations $%
\pm 5\%$, $\pm 3\%$, and $\pm 1\%$ give rise to the stability border at $%
\eta _{\max }^{2}=1.02$, $1.055$, and $1.08$, respectively, while Eq. (\ref%
{crit}) yields $\eta _{\max }^{2}\approx 1.15$ in the same case.

\begin{figure}[tbh]
\centerline{%
\includegraphics[width=7.0cm]{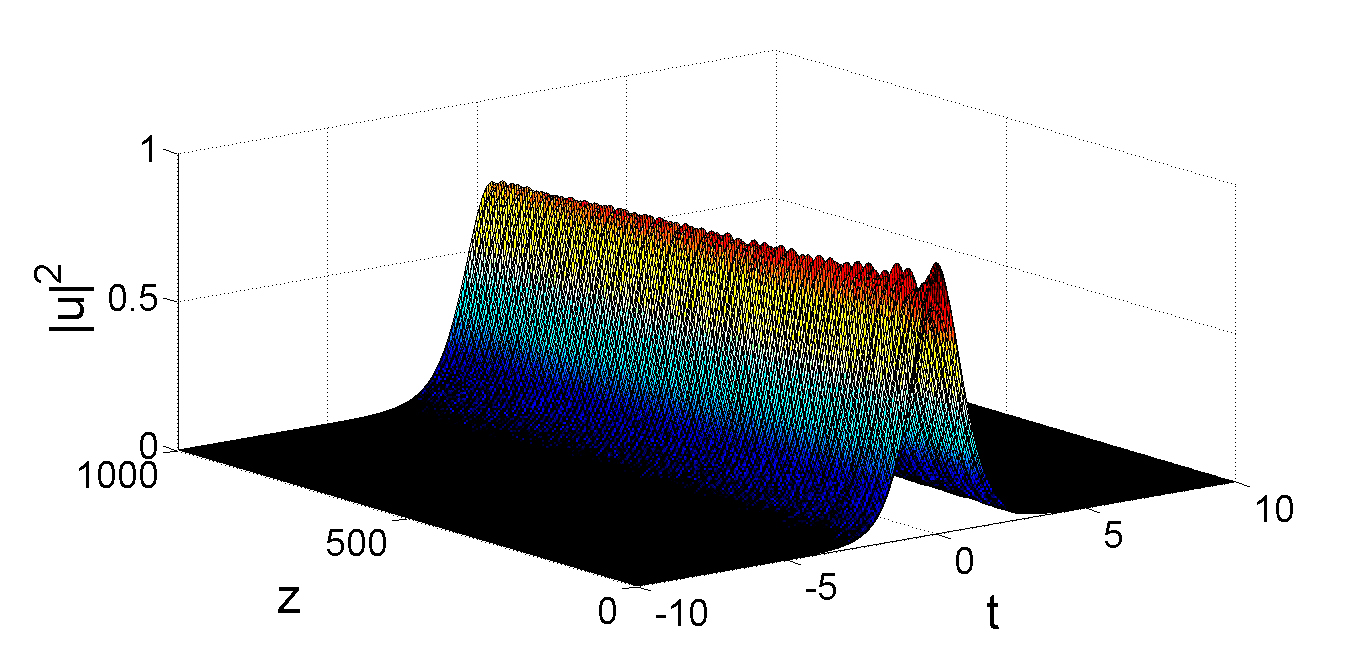}}%
\centerline{%
\includegraphics[width=7.0cm]{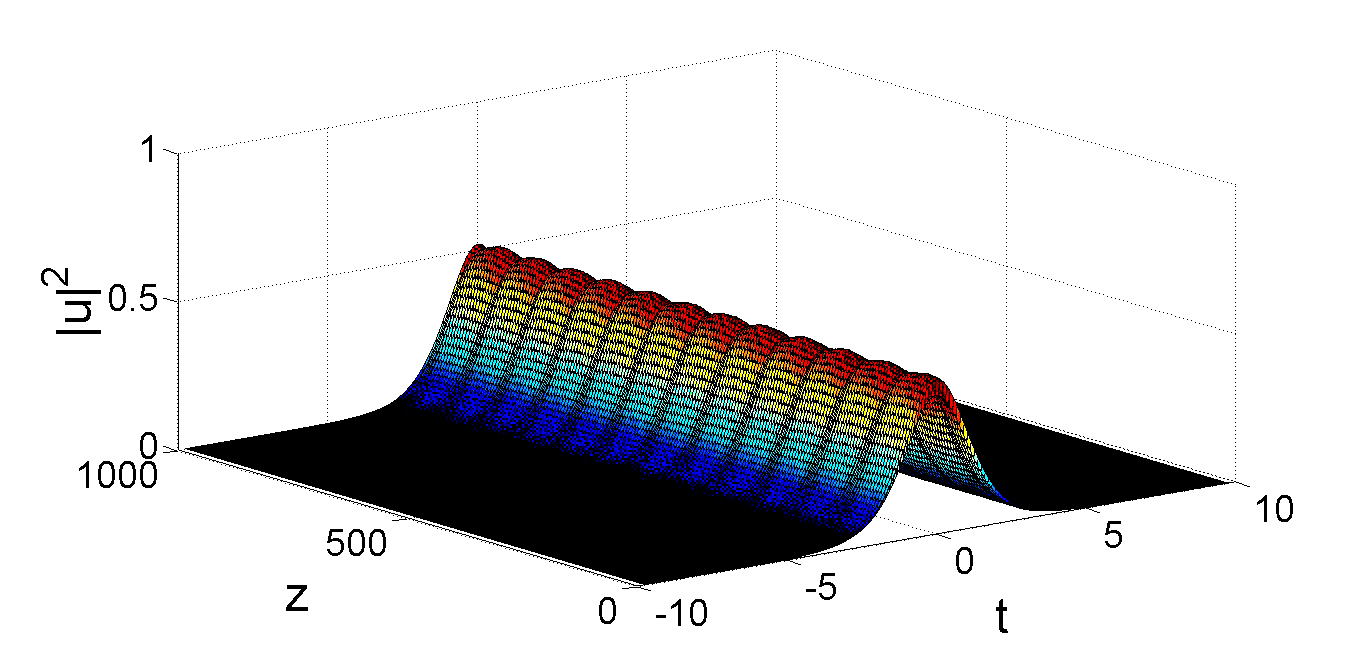}}
\caption{(Color online) Examples of the relaxation of stable perturbed $%
\mathcal{PT}$-symmetric (a) and antisymmetric (b) solitons, with $\protect%
\gamma =0.5,\protect\eta =0.9$ and $\protect\gamma =0.5,\protect\eta =0.7$,
respectively. Here \ and in the case shown in Fig. \protect\ref{fig3}, the
evolution of the $v$-component is similar to what is shown for $u(z,t)$.}
\label{fig2}
\end{figure}

Examples of the stable propagation of perturbed solitons of both types are
displayed in Fig. \ref{fig2}. These examples are shown for the solitons
located close to the stability border, cf. Fig. \ref{fig1} [in particular,
Eq. (\ref{crit}) yields $\eta _{\mathrm{cr}}(\gamma =0.5)\approx 1.07$, cf. $%
\eta =0.9$ in Fig. \ref{fig2}(a)]. Unlike the dual-core coupler \cite{Herb}-%
\cite{Pak}, the $\mathcal{PT}$ system cannot support asymmetric solitons, as
the balance between the gain and loss is impossible for them. Accordingly,
the instability of solitons with $\eta >\eta _{\max }$ leads to a blowup of
field $u$ and decay of $v$ in the direct simulations (not shown here).

\begin{figure}[tbh]
\centerline{\includegraphics[width=8.0cm]{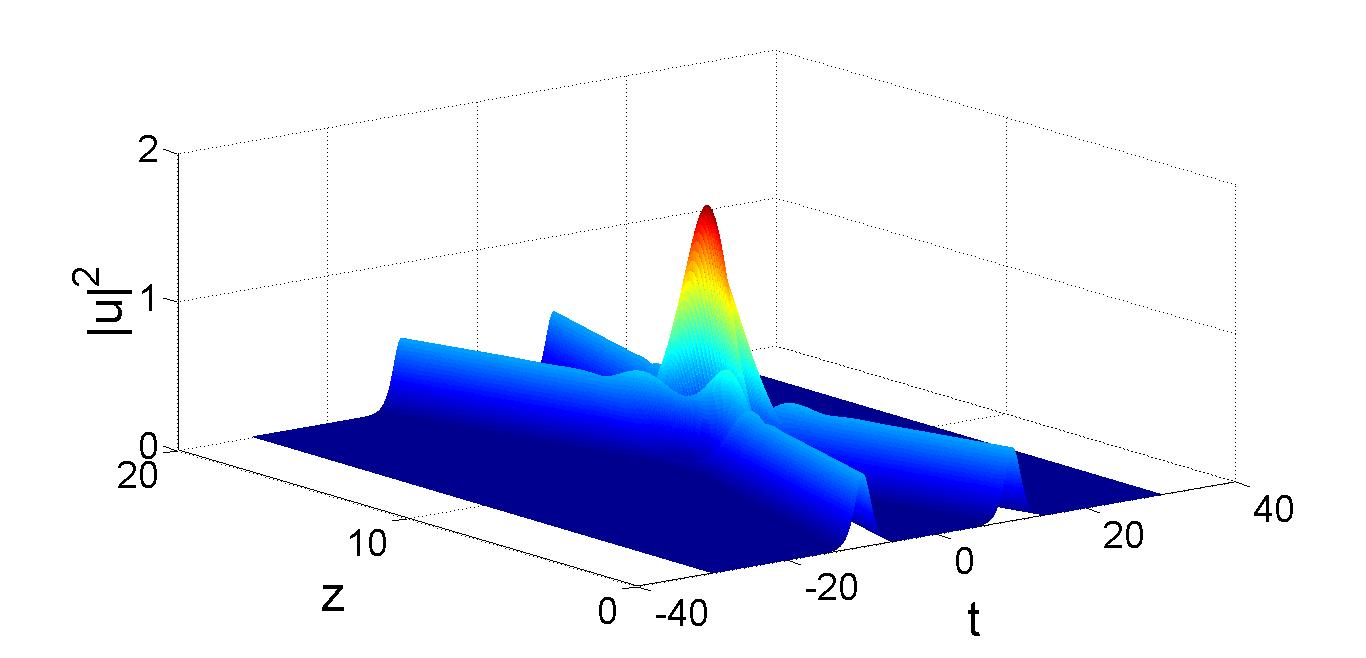}}
\caption{(Color online) The elastic collision between stable solitons of the
two different types, with $\protect\eta =0.7$, boosted by frequency shift $%
\protect\chi =\pm 5$ at $\protect\gamma =0.7$.}
\label{fig3}
\end{figure}

The Galilean invariance of Eqs. (\ref{uv}) suggests to consider collisions
between stable solitons, setting them in motion by means of \textit{boosting}%
, $\left\{ u,v\right\} \rightarrow \left\{ u,v\right\} \exp (\pm i\chi t$),
with frequency shift $\chi $. The results is that the collisions always seem
elastic. A typical collision between $\mathcal{PT}$-symmetric and
antisymmetric solitons is displayed in Fig. \ref{fig3}.

In the case of the ``supersymmetry", $\gamma =1$, the stability
regions in Fig. \ref{fig1} shrink to nil. In this case, the
linearization of Eqs. (\ref{uv}) with $\gamma =\Gamma =\kappa $
around NLS
solution (\ref{U}) leads to the following equations for perturbations $%
\delta u$ and $\delta v$:
\begin{equation}
\hat{L}\left( \delta u+i\delta v\right) =0,\hat{L}\left( \delta u-i\delta
v\right) =-2i\kappa \left( \delta u+i\delta v\right) ,  \label{L}
\end{equation}%
where $\hat{L}\delta u\equiv \left[ i\partial _{z}+(1/2)\partial
_{tt}+2|U|^{2}\right] \delta u+U^{2}\delta u^{\ast }$ defines the
linearization operator of the NLS equation. If the NLS solution is stable,
the first equation (\ref{L}) produces no instability, while the second gives
rise to the \textit{resonance}, as $\left( \delta u+i\delta v\right) $ is a
zero mode of entire operator $\hat{L}$, which includes term $i\partial _{z}$%
. According to the linear-resonance theory, the respective perturbation $%
\left( \delta u-i\delta v\right) $ will be unstable, growing $\sim z$
(rather than exponentially). Direct simulations of Eqs. (\ref{uv}) with $%
\gamma =\kappa $ confirm that the solitons are unstable, the character of
the instability being consistent with its subexponential character.

The ``supersymmetric" solitons may be stabilized by means of the
\textit{management} technique \cite{Radik,book}, periodically
reversing the common sign of $\gamma =\Gamma =\kappa $, which does
not break the supersymmetry of system (\ref{uv}). Flipping $\gamma $
and $\Gamma $ means switch of the gain between the two cores, which
is possible if both are doped \cite{doping}. The coupling
coefficient, $\kappa $, cannot flip by itself, but the signal in one
core may pass $\pi $-shifting plates, which is tantamount to the
periodic sign reversal of $\kappa $.

\begin{figure}[tbh]
\centerline{\includegraphics[width=8.0cm]{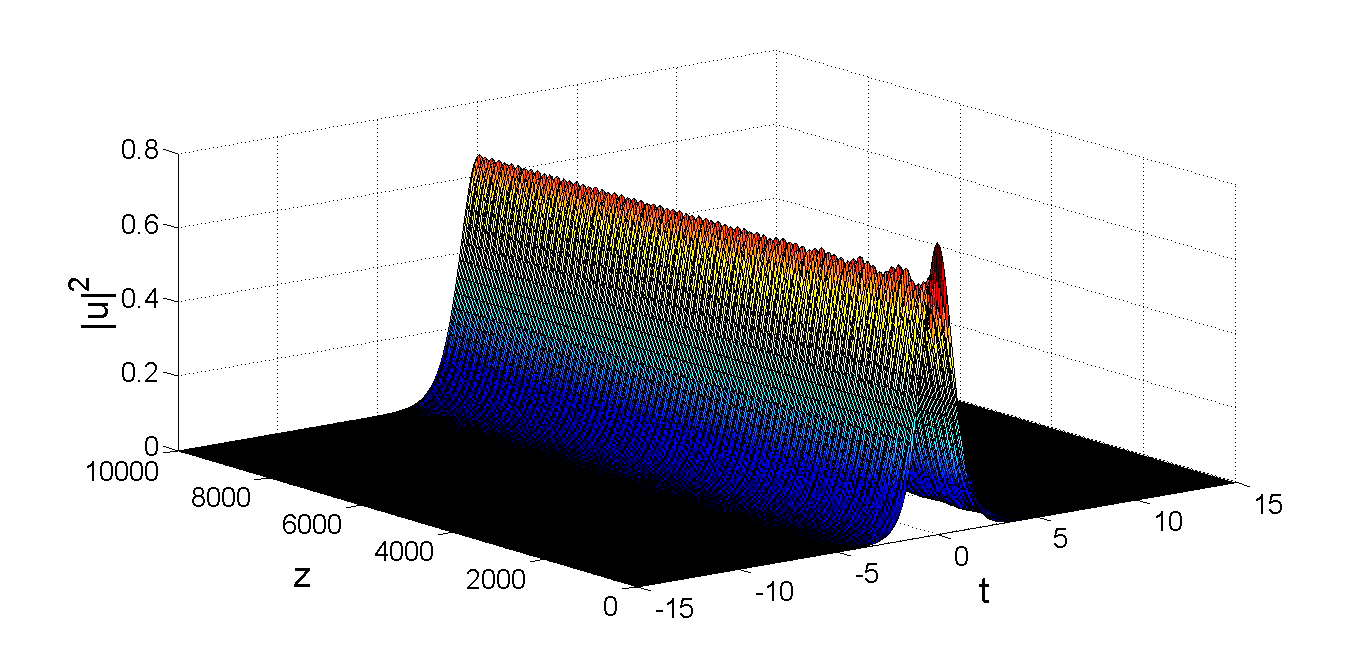}}
\caption{(Color online) Self-trapping of the soliton in the
``supersymmetric" system ($\Gamma =\protect\gamma =\protect%
\kappa =1$) subject to the management with period $\Delta z=2$. The input is
$u_{0}=-iv_{0}=0.9\mathrm{sech}(t)$.}
\label{fig4}
\end{figure}

Ansatz (\ref{U}) still yields exact solutions to Eqs. (\ref{uv}) with
coefficients $\Gamma =\gamma =\kappa $ subjected to the periodic management.
On the other hand, the replacement of $\kappa $ by the periodically flipping
coefficient destroys the resonance in Eq. (\ref{L}). In simulations, this
management mode indeed provides for self-trapping of robust solitons from
inputs significantly different from the exact soliton solution. An example
is shown in Fig. \ref{fig4} for the input with a relatively large
perturbation, in the form of a $10\%$ amplitude deviation from the exact
solution. A detailed analysis of the management will be reported elsewhere.

In conclusion, we have found the families of exact symmetric and
antisymmetric solitons in the $\mathcal{PT}$-symmetric dual-core system with
the Kerr nonlinearity. For the symmetric family, the stability border (\ref%
{crit}) was found in the analytical form, and corroborated by
simulations. The stability region for the antisymmetric solitons was
obtained in the numerical form. Collisions between solitons of both
types are elastic. In the ``supersymmetric" limit, with the equal
gain, loss and inter-core-coupling coefficients, the two soliton
species merge into one, which is subject to the subexponential
instability, and can be stabilized by means of the periodic
management. It may be interesting to extend the spatial-domain model
by including a periodic linear or nonlinear potential, which may
give rise to other soliton modes \cite{Barcelona}, and to include
the intermode dispersion in the temporal domain \cite{Chiang}.

\textbf{References with titles}

\textbf{References without titles}

\end{document}